\documentstyle[emulateapj,apjfonts]{article}
\slugcomment{Received: 13 March 2000, Accepted: 27 March 2000: ApJ}
\lefthead{Yong et al.}
\righthead{Mass Loss On the Horizontal Branch}
 
\begin{document}

\title{Mass Loss On the Horizontal Branch: an Application to NGC\,6791}
\author{Hwanjong Yong and Pierre Demarque}
\affil{Department of Astronomy, Yale University, P.O. Box 208101, New Haven, CT 06520-8101\\yong@astro.yale.edu, demarque@astro.yale.edu}
\author{Sukyoung Yi}
\affil{Center for Space Astrophysics, Yonsei University, Seoul 120-749, Korea, and California Institute of Technology, Pasadena, CA 91125\\ 
yi@srl.caltech.edu}
\authoremail{yi@srl.caltech.edu}

\begin{abstract}

	The presence of a substantial number of hot stars in the {\it 
extremely metal-rich} open cluster NGC\,6791 has been a mystery.
	If these hot stars are in their core helium burning phase,
they are significantly bluer (hotter) than predicted by canonical
stellar evolution theory.
	No obvious explanation is available yet.
	We consider the effects of mass loss {\em during the evolution 
of horizontal branch} (HB) stars as their possible origin. 
	We find that the addition of mass loss causes HB stars to evolve to be 
{\it hotter} and {\it fainter}.
	Mass loss has a more pronounced effect for less massive stars and
thus naturally widens the temperature (and color) distribution on the HB.
	If mass loss rates are higher for more metal-rich stars,
this phenomenon would be even more pronounced in the metal-rich populations,
such as NGC\,6791.
	We find that mass loss on the HB may be a viable method of forming 
subdwarf B (sdB) stars both in the field and in clusters, especially when the
metallicity is high.

\end{abstract}


\section{Introduction}

	Understanding the origin and evolution of hot horizontal branch (HB)
stars would have a significant impact on several areas of astronomy.  
	For example, it is important in the study of globular cluster 
HB morphology, which constrains the formation history of the Galaxy 
(Searle \& Zinn 1978).  
	The global picture of HB morphology appears to be well understood.
	HB morphology is most affected by the metallicity of the cluster: 
more metal rich clusters exhibit redder HBs (``the first parameter'' of
HB morphology).  
	However, observations in some globular clusters show HBs which exhibit 
color distributions not expected for their metallicities.  
	To explain this deviation, a ``second parameter'' is invoked. 

	Age has been the most successful global candidate for the
"second parameter" (Searle \& Zinn 1978; 
Lee, Demarque, \& Zinn 1994; for recent review, see 
Stetson, VandenBerg, \& Bolte 1996; Sarajedini, Chaboyer, \& Demarque 1997).
	However, recent observations show an increasing number of HB 
morphologies with anomalous features, e.g. bimodality in the color 
(temperature) distribution, extended blue tails, and gaps 
(e.g., Sosin et al. 1997; Rich et al. 1997; Catelan et al. 1998). 
	Most of these anomalies are found near the blue end of the HB. 
	Within the framework of the canonical theory of single star 
evolution, they appear to be inconsistent with a scenario where metallicity 
and age are the only two parameters that govern the HB morphology.  

	NGC\,6791 is an extreme example that exhibits deviations from the 
first and second parameter phenomena. 
	This open cluster is believed to be extremely metal-rich, 
[Fe/H] $= 0.2$ -- 0.5 (Friel \& Janes 1993; Kaluzny \& Rucinski 
1995; Peterson \& Green 1998), which would suggest a predominantly red HB 
clump, following the first parameter phenomenon.
	Yet, nearly a third of its helium burning population (HB stars) are 
extremely blue (Liebert et al. 1994; Green et al. 1997) with a large 
temperature separation from the red clump.
	Dynamical effects do not seem to provide a promising explanation
for this anomaly,
because stellar densities are small in open clusters.

	Figure 1 shows the observed color-magnitude diagram (CMD) and a 
synthetic HB model based on canonical stellar models.
	The observed CMD of NGC\,6791, which is from Kaluzny \& Rucinski 
(1995), shows approximately 23 red clump stars and 9 hot stars.
	These hot stars have also been identified by ultraviolet (UV)
observations using the {\it Ultraviolet Imaging Telescope} 
(Landsman et al. 1998).
	The synthetic HB (diamonds) is based on the arbitrarily chosen
chemical composition, i.e. [Fe/H]$ = 0.33$ and $Y=0.31$
(equivalent to ${\Delta Y / \Delta Z = 1.5}$).
	For illustrative purposes, a 9 Gyr isochrone
from the main sequence (MS) to the red giant branch (RGB), is shown
on the same plot. 
	For $E(B-V) =0.125$ and $m-M=13.35$, the match 
is reasonable.
	It is beyond the scope of this paper to derive the age of this
cluster or any other parameters by matching the CMD with isochrones.

	For the synthetic HB construction, we have assumed 
a Gaussian mass dispersion parameter
$\sigma_{HB} = 0.04 M_\odot$,  and a Reimers mass loss parameter 
$\eta = 1.0$ (Lamers \& Cassinelli 1999; 
Willson, Bowen, \& Struck 1996).
	Readers are referred to Yi, Demarque, \& Oemler (1997b) for
details of the synthetic HB construction.
	We assume that there are approximately 32 HB stars in NGC\,6791,
and we have added some random scatter in both $M_{V}$ and $B-V$. 
	The model reproduces, in accordance with the first 
parameter phenomenon, the prominent red clump seen in the data, 
but it lacks hot stars.

	NGC\,6791 is the oldest Galactic open cluster known to-date at 
6.5-9 Gyr (Demarque, Green, \& Guenther 1992; Phelps et al. 1994;
Garvanich et al. 1994; Chaboyer, Green, \& Liebert 1999). 
	The most recent estimate from Chaboyer et al. (1999) suggests
$8.0 \pm 0.5$ Gyr for the age of this cluster.
	However, even at the largest age estimate so far (9 Gyr), a 
synthetic HB based on the conventional input parameters would not reproduce 
anything even close to the HB morphology of NGC\,6791.
	By increasing its age by a couple of Gyrs, we could produce some hot 
HB stars, but their numbers would still be much smaller than observed.
	Thus, the age effect alone seems unsuccessful in explaining the
HB color distribution.

	A relatively large value of the dispersion 
($\sigma = 0.06 M_{\odot}$, in 
comparison to the conventional value 0.03 -- 0.04) in the Gaussian mass 
distribution on the HB would result in a larger fraction of hot HB stars.
	This may match the blue-to-red HB star ratio found in NGC\,6791 
better, but the strong color-bimodality in NGC\,6791 would not be reproduced.

	One or more local parameters must be at work in this cluster.
	As candidate mechanisms that produce hot HB stars, a few additional  
scenarios have been proposed lately. 
	Sweigart (1997) has suggested that helium mixing on the red giant 
branch would cause further evolution beyond the tip, resulting in higher mass 
loss on the giant branch.  
	As the color dispersion on the HB is the result of varying
envelope masses, the enhanced mass loss results in a bluer HB.  
	D'Cruz et al. (1997) assumed an episode of extreme mass 
loss at the tip of the 
giant branch to explain blue HB stars.

\placefigure{fig1}
\parbox{3.2in}{\epsfxsize=3.2in \epsfbox{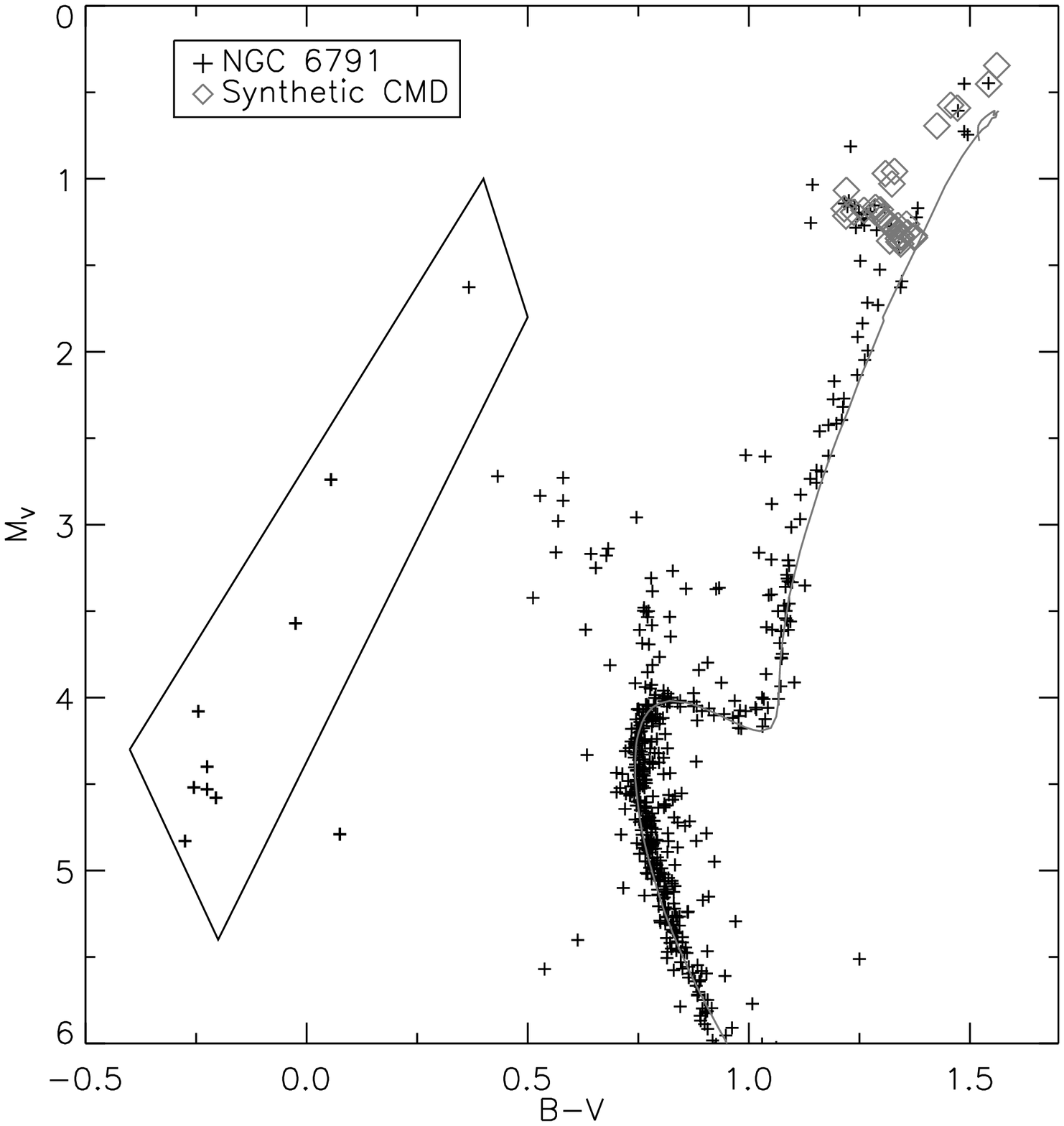}}
\centerline{\parbox{3.2in}{\small {\sc Fig. 1}
Observed CMD of NGC\,6791 and a synthetic HB model. Note the presence of 
hot HB stars (enclosed in a box) only in the observed CMD.
}}
\vspace{0.2in}
\addtocounter{figure}{1}

	In this letter, we explore yet another possible formation scenario 
for blue HB stars: mass loss during the core helium burning phase.
	Demarque \& Eder (1985) first investigated this effect for low 
metallicity HB stars, and we extend on their work here. 
	We will show that the inclusion of mass loss on the HB works in 
the direction of explaining the color distribution of the HB stars
in NGC\,6791.

\section{Evolutionary Model Construction}

	We used the Yale stellar evolution code (YREC) for the construction
of HB evolutionary models.
	The detail of physics are the same as in Guenther et al. (1992) with 
the addition of updated opacities.  
	For log $T \ge 4.00$, OPAL opacities are used 
(Iglesias \& Rogers 1996), and Alexander low temperature opacities are
used for log $T < 4.00$ (Alexander \& Ferguson 1994).  
	The ratio $(\alpha)$ of the mixing length to pressure scale 
height in the convection zone, was assumed to be 1.9, which 
is the value used in a standard solar model with the same input physics 
(Guenther et al. 1992). 
	This parameter only affects the radii of the very coolest models, and
has little impact on our conclusions.  
	Although He diffusion and complex heavy element segregation processes 
are known to take place in the atmospheres of sdB and sdO stars (Michaud,
Vauclair, \& Vauclair 1983; Michaud et al. 1985; Unglaub \& Bues 1998), 
their effect to the internal structure and radius calculation is negligible. 

	In addition, a mass-loss routine was included into YREC.  
	As we propose no mechanism for mass loss, and any empirical mass loss 
rate, such as Reimers mass loss (which is based on red supergiant 
observations), does not apply here, the simplest approach was taken.
	The mass loss rate, which is supplied by the user, is assumed to 
be constant over time, and is used to determine the amount of mass loss for a 
given time step.  
	This mass is then removed from the envelope, which is taken here to 
be the region of the hydrogen burning shell and outward, at each time step.
	The model is then allowed to evolve with its new mass.  
	The energy needed to eject the mass against the gravitational field 
is also calculated and removed from the envelope (see also, 
Koopmann et al. 1994 and Demarque \& Eder 1985).

	While mass loss on the HB is an {\it ad hoc} assumption, 
it is possible to justify at least some mass loss. 
	Mass loss is directly observed in the Sun at a rate of 
${10^{-14} M_{\odot} yr^{-1}}$.  
	It has also been pointed out that mass loss should be enhanced in 
pulsating stars (Willson 1988).  
	As RR Lyrae stars are pulsators on the horizontal branch, one should 
expect some mass loss in these stars.  
	There is also both observational and theoretical evidence for 
pulsation in hot horizontal branch stars.
	Recent observations show that sdB stars pulsate as well (Kilkenny 
et al. 1998 and references therein).  
	This pulsation was predicted theoretically by Charpinet et al. 
(1997a, 1997b).
	The effect of mass loss in HB stars while they evolve through
the instability strip was studied by Koopmann et al. (1994). 
	They found that the mass loss rate has an upper limit of  
${10^{-9} M_{\odot}yr^{-1}}$.
	A higher mass loss rate would break the good agreement between 
synthetic and observed HBs in the globular cluster M4.
	This provides an upper limit for the mass loss rate on the HB 
in the intermediate temperature (RR Lyrae) region.  

	HB stars hotter than RR Lyrae variables may be in 
conditions under which larger mass loss may take place.
	According to Abbott's formula (1982), the mass loss rate for these 
stars for the solar composition should be around 
2 -- 4 ${10^{-12} M_{\odot} yr^{-1}}$.  
	We can think of two effects which should increase this number. 
	The first is the fact that the sdB stars pulsate.
	Second is that the mass loss rate should scale with the 
metallicity (Abbott 1982; Willson et al. 1996; Lamers \& Cassinelli 1999), 
so the assumption of higher mass loss rate in more metal rich stars, such 
as the blue HB stars in NGC\,6791, could be justified.  
	Although it would be very difficult to observe directly 
such low mass loss rates in HB stars, there is spectroscopic evidence which 
supports the presence of mass loss in sdB and sdO atmospheres. 
	The abundance anomalies observed in the atmospheres of sdB and sdO 
stars, and their lifetimes, can only be explained when the interaction of 
both diffusion and mass loss are taken into account (Michaud et al. 1983, 
1985; Unglaub \& Bues 1998 for the sdO stars).  
	More detailed spectroscopy and atmospheric modeling in the future will
reveal the time dependence and intensity of these winds.

	The mass loss rates used for the present study are in the range 
${10^{-10} \le \dot M (M_{\odot} yr^{-1}}) \le 10^{-9}$.
	Models have been constructed for the metallicity range of 
${0.004 \le Z \le 0.1}$ and approximately for ${\Delta Y /\Delta Z = 1.5}$
and 3.  

	The isochrones for the hydrogen burning phases 
that we have used in this paper are the Sukyoung Yi
version of the Yale Isochrones (Yi, Demarque, \& Oemler 1997b; available
from http://shemesh.gsfc.nasa.gov).
	The input physics used in these isochrones are somewhat older
than the ones used in our HB models, but they are qualitatively
compatible to our HB models.

\section{Results and Discussion}

	In order to study the possibility of forming blue HB stars through mass
loss during the HB phase of stellar evolution, we have constructed 
evolutionary tracks for HB stars of varying metallicity and envelope mass.  
	Figure 2 shows HB evolutionary tracks and 6 hot stars in NGC\,6791
whose temperatures and luminosities have been measured by Landsman et al.
(1998) based on their UV fluxes.
	The track furthest to the right is the evolutionary track without 
mass loss and each successive track to the left has a larger mass loss rate;
${10^{-10}}$, ${10^{-9.5}}$, ${10^{-9} M_{\odot} yr^{-1}}$, respectively.
	The effect of mass loss shown in Figure 2 is representative of the 
effects of mass loss in the other models calculated.  
	Several points can be made by examining Figure 2.  
	First, the greater the mass loss rate assumed, the hotter the star 
becomes.
	This is not a surprising result. 
	It is generally believed that the HB is a sequence of stars which have 
the same core mass and varying envelope masses.
	The smaller the envelope mass, the hotter the star appears, as a 
deeper and hotter region is exposed with a smaller envelope.  
	It seems natural then that mass loss during the HB phase will produce
hotter stars with smaller envelopes.  

\placefigure{fig2}
\parbox{3.2in}{\epsfxsize=3.2in \epsfbox{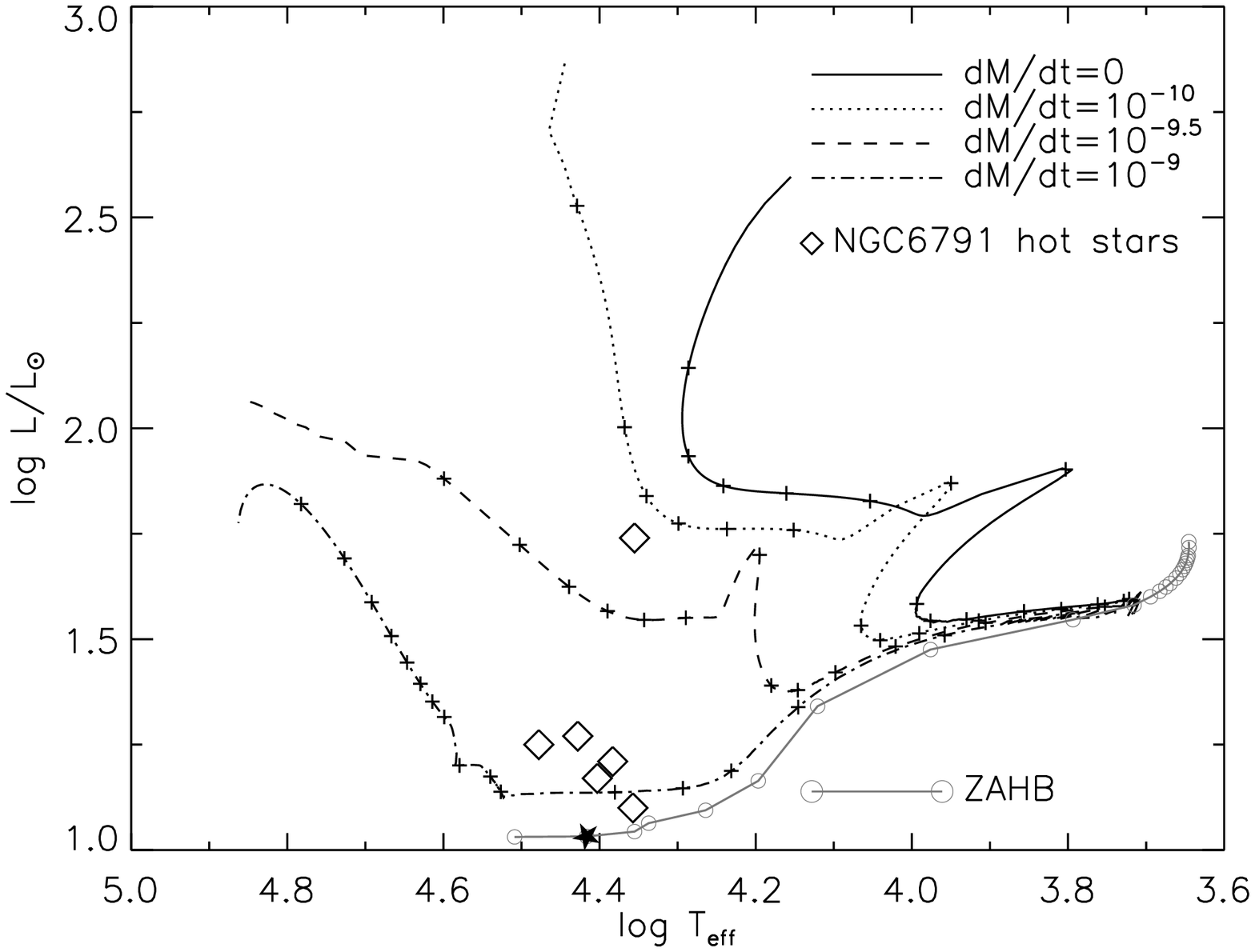}}
\centerline{\parbox{3.2in}{\small {\sc Fig. 2}
Evolutionary tracks for stars starting on the zero-age HB for Z=0.04, 
${\Delta Y /\Delta Z = 3}$ and envelope mass of ${0.07M_\odot}$.  
The tracks of ${\Delta Y /\Delta Z = 1.5}$ show
qualitatively the same phenomenon. 
The zero-age HB locus, which is insensitive to the adopted mass loss rate, is 
shown near the bottom of the plot. Tracks with increasing mass loss
rates are shifted toward the hotter and fainter regions of the H-R diagram.  
The crosses indicate time intervals of ${10^7}$ years.
Also shown are the hot stars in NGC\,6791 whose temperatures and luminosities
have been measured from their UV fluxes (Landsman et al. 1998).
}}
\vspace{0.2in}
\addtocounter{figure}{2}

	At the same time, a larger mass loss rate also results
in the stars becoming fainter than the stars without mass loss,
particularly for the larger mass loss rates.
	As our mass loss scheme removes mass from the envelope, which is 
defined to be the hydrogen burning shell and outwards, the total amount of 
hydrogen burning is reduced.  
	This results in the stars evolving with smaller luminosities.  
	Figure 2 shows that the atmospheric properties of the hot
stars in NGC\,6791 (diamonds) are well matched by the theoretical
models with mass loss if their envelope masses are approximately 
$0.07 M_\odot$.
	If there is no mass loss on the HB, their properties can be
matched only if their envelope masses are as small as $0.001 M_\odot$
(the filled star symbol at log\,$T_{\rm eff} \approx 4.4$ and
log\,L/L$_{\odot} \approx 1.1$).
	Whether HB stars can be born with such low envelope masses at all
has been questioned by many groups: ``the minimum envelope mass hypothesis'' 
(see the discussion in \S3.2 of Yi et al. 1997b).
	With mass loss on the HB, one can match the observed data
without violating the minimum envelope mass hypothesis.	

\placefigure{fig3}
\parbox{3.2in}{\epsfxsize=3.2in \epsfbox{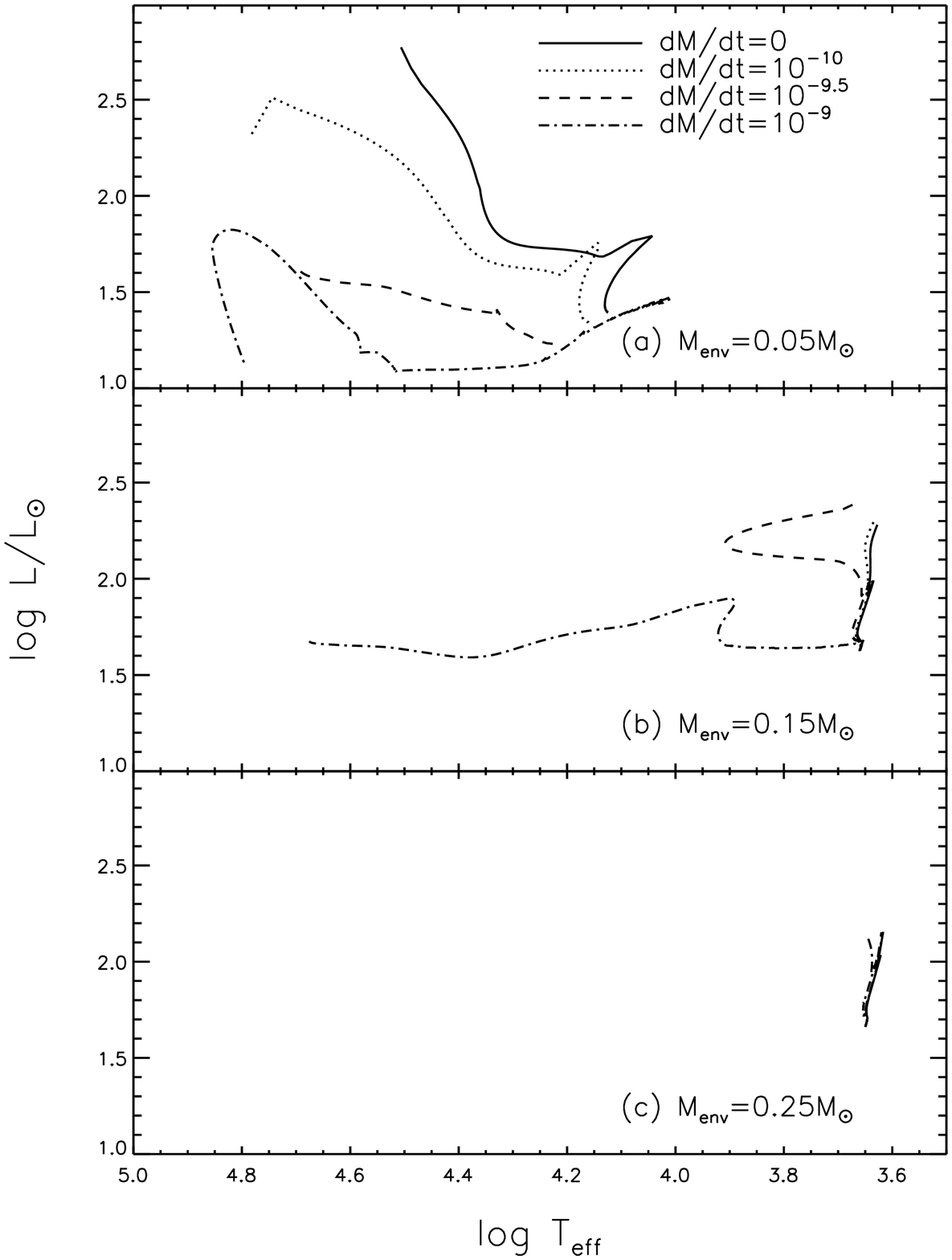}}
\centerline{\parbox{3.2in}{\small {\sc Fig. 3}
Comparison of evolutionary tracks for a constant metallicity Z=0.04
and varying envelope masses.  Mass loss has a more profound effect on low
envelope mass stars.
}}
\vspace{0.2in}
\addtocounter{figure}{3}

	It should be noted that the adoption of mass loss on the HB does 
not change the zero-age HB locus at all in our scenario, because we 
are only comparing the tracks with and without mass loss when they 
begin their HB evolution at the same location in the CMD.

	The inclusion of mass loss on the HB also contributes to 
the wide color distribution in the HB of NGC\,6791.
	Figure 3 shows evolutionary tracks for HB stars with varying envelope
masses.  
	Each panel is for a given envelope mass with four mass
loss rates plotted for each.  
	The composition for this particular set of models is $Z=0.04$ and 
${\Delta Y / \Delta Z = 3}$.
	Mass loss has little effect to the evolution of the HB stars with 
large envelope masses but significantly affects that of less massive stars.
	This mass dependency would generate a wider color distribution in 
the HB.
	It may even create a color bimodality (or multimodality), as seen
in the hot ends of the HBs of several globular clusters, if the mass 
dependency is not a smooth function of mass.  
	This may explain the lack of intermediate HB stars in NGC\,6791.  
	A more detailed modeling will show if this is a plausible
explanation to the case of NGC\,6791.

	The effects of such mass loss in the HB phase have an impact on the
study of ultraviolet spectral evolution of old stellar populations as well.
	Numerical studies by Demarque and Pinsonneault (1988) 
and by Horch, Demarque, \& Pinsonneault (1992) suggested that in the advanced 
evolution of metal rich HB stars, stars with small envelopes evolve to become
UV bright objects, the so-called slow blue phase (SBP) stars, rather than 
evolving into asymptotic giant branch stars (c.f. the AGB-manqu\'{e} stars in
Greggio \& Renzini 1990).
	Yi, Demarque \& Kim (1997a) elaborated on the theory 
and found a theoretical explanation to the SBP phenomenon and to its
metallicity dependence, that is, the transitional mass (the maximum mass 
of a star that becomes a SBP star) increasing with metallicity.  
	According to this theory, more metal-rich stars become UV bright SBP 
stars more easily.
 
	With the additional mass loss on the HB that we consider in this 
paper, stars would become SBP stars even more easily. 
	Moreover, if mass loss on the HB increases with increasing metallicity
and increasing effective temperature, as various studies suggest, it would 
cause metal-rich stellar populations to develop UV bright stars more quickly.
	Because SBP stars are considered important UV sources in elliptical 
galaxies (e.g., Tantalo et al. 1996; Yi et al. 1997a; O'Connell 1999) 
and the UV flux of elliptical galaxies may constrain the ages of galaxies 
(Yi et al. 1999), this study has an impact on extragalactic astronomy. 
	Landsman et al. (1998) already pointed out that a population like
NGC\,6791 would exhibit a UV upturn with the magnitude shown in giant 
elliptical galaxies.
	If mass loss on the HB indeed contributes to the production of
hot stars in the metal-rich environments, our current age estimates
of giant elliptical galaxies are likely to be systematically overestimated.

	We should point out that there is no direct observational evidence for 
mass loss as substantial as we have assumed to occur on the HB.  
	In fact, besides Koopmann et al.'s upper limit on the
mass loss rate on the HB (${10^{-9} M_{\odot}yr^{-1}}$), the diffusion
study of Michaud et al. (1985) set a more strict upper limit of
${10^{-14} M_{\odot}yr^{-1}}$.
	They suggest that if mass loss rate is larger than this it would
be difficult to reproduce the silicon underabundance observed in sdB stars.
	If this is true, the mass loss rates investigated in our study
may be too large to justify, unless mass loss rate increases substantially
in the metal-rich regime.
	Once again, we do not advocate any value or mechanism for the
mass loss on the HB.
	Instead, we wanted to see what level of mass loss rate on the HB
would be needed if it is the only process that is responsible for the 
production of hot HB stars.
	Based on a cursory inspection of our new HB tracks with mass loss,
adoption of such low mass rates ($ < {10^{-14} M_{\odot}yr^{-1}}$)
would not make any appreciable difference in the HB morphology.

	Some of the blue HB stars have been observed to be spectroscopic 
binaries (Green et al. 1997).  
	This brings up the possibility that the blue HB stars are
formed through binary evolution.  
	Some of the formation scenarios such as helium white dwarf mergers 
(Bailyn \& Iben 1989, Iben 1990) can perhaps be ruled out based on the 
fact that the luminosities are in a small range (see Figure 2).  
	This small range of the luminosities of the hot stars is also
in conflict with the binary scenario (Landsman et al. 1998).
	Mengel, Norris, \& Gross (1976) demonstrated that blue HB stars 
can form through binary evolution with mass transfer via Roche Lobe overflow.  
	However, the conditions under which this occurs are probably too 
infrequently met, although further study, applicable to metal rich stellar
populations, is needed.
	Furthermore, it is unclear whether all blue HB stars in NGC\,6791 
are in mass-exchanging binary systems.
	Therefore, a mechanism to form blue HB stars through single star 
evolution is still needed.  
	Mass loss on the HB may be such a mechanism.

\acknowledgements

        This work was supported in part by the Creative Research Initiative 
Program of the Korean Ministry of Science \& Technology grant (S.Y.), and 
grant NAG5-8406 from NASA (P.D.).

\end{document}